\documentclass[10pt,journal]{IEEEtran}

\usepackage{cite}
\usepackage{amsmath,amssymb,amsfonts}
\usepackage{algorithm}
\usepackage{algorithmic}
\usepackage{graphicx}
\usepackage{textcomp}
\usepackage{xcolor}
\usepackage{booktabs}
\usepackage{multirow}
\usepackage{array}
\usepackage{tabularx}
\usepackage{makecell}
\usepackage{url}
\usepackage{balance}
\usepackage{float}
\usepackage{placeins}

\newcommand{\ket}[1]{\left|#1\right\rangle}
\newcommand{\bra}[1]{\left\langle#1\right|}
\newcommand{\proj}[1]{\ket{#1}\!\bra{#1}}
\newcommand{\Tr}{\operatorname{Tr}}
\begin{document}
\bstctlcite{IEEEexample:BSTcontrol}

\title{A CPU+DCU Heterogeneous Parallel Framework for Post-Processing Reconstruction in Quantum Circuit Cutting}

\author{%
Qingqing~Jiang,
Weidong~Liu,
Yufu~Liu,
Ruiqing~He,
Jiandong~Shang,
Hengliang~Guo,
and Qiang~Chen%
\IEEEcompsocitemizethanks{%
\IEEEcompsocthanksitem
Qingqing Jiang, Weidong Liu, Yufu Liu, Jiandong Shang,
Hengliang Guo, and Qiang Chen are with the
National Supercomputing Center in Zhengzhou,
Zhengzhou University, Zhengzhou 450001, China.
\IEEEcompsocthanksitem
Qingqing Jiang, Weidong Liu, and Yufu Liu are also with the
School of Computer and Artificial Intelligence,
Zhengzhou University, Zhengzhou 450001, China.
\IEEEcompsocthanksitem
Ruiqing He is with the
School of Communication and Artificial Intelligence
and the School of Integrated Circuits,
Nanjing Institute of Technology,
Nanjing 211167, China.
\IEEEcompsocthanksitem
Corresponding author: Qiang Chen
(e-mail: qiangchen@zzu.edu.cn).
}%
}

\markboth{IEEE Transactions on Parallel and Distributed Systems,~Vol.~xx, No.~x, Month~2026}%
{Author \MakeLowercase{\textit{et al.}}: A CPU+DCU Heterogeneous Parallel Framework for Post-Processing Reconstruction in Quantum Circuit Cutting}

\IEEEtitleabstractindextext{%
\begin{abstract}
In the NISQ era, limited qubit resources make it
difficult to execute large quantum circuits directly on real hardware. Quantum
circuit cutting mitigates this limitation by decomposing a large circuit into
smaller subcircuits, but it shifts substantial overhead to classical
post-processing. As circuit size, complexity, and cut count increase,
reconstruction becomes a major computational and storage bottleneck.
This paper presents a CPU+DCU heterogeneous parallel framework for
circuit-cutting post-processing reconstruction. Instead of constructing a dense
\(2^n\)-dimensional probability vector or returning only high-probability
states, the framework reconstructs the nonzero-probability states in the original output distribution from subcircuit measurement results. It combines heterogeneous CPU+DCU execution with a high/low-word integer representation for global basis-state indices beyond 64 bits and a three-level cooperative storage mechanism spanning device memory, host memory, and out-of-core storage.
Experiments on the Songshan supercomputer show that the framework maintains
high reconstruction fidelity while achieving up to \(259\times\) speedup over an optimized serial baseline on linear-cluster states and up to \(4\times\) speedup over a homogeneous
CPU-parallel method on random circuits. The framework can also complete reconstruction tasks at the hundred-qubit scale. These results demonstrate that HPC-oriented heterogeneous reconstruction can effectively alleviate the classical post-processing bottleneck and improve reconstruction scalability.
\end{abstract}

}

\maketitle
\IEEEdisplaynontitleabstractindextext

\section{Introduction}

\IEEEPARstart{Q}{uantum} computing is widely regarded as a promising
paradigm for achieving computational advantages in machine learning\cite{WOS:000410555900032}, quantum
simulation\cite{WOS:A1982NR25300003,
WOS:A1996VD42800029},  chemistry\cite{WOS:000273820700014,WOS:A1997XZ11400050}, and other scientific fields\cite{montanaro2016quantum}. Representative quantum
algorithms have demonstrated exponential speedups for integer factorization\cite{shor1994algorithms}
and polynomial speedups for unstructured search\cite{grover1996fast}. In practice, however, current
quantum processors remain constrained by the noisy intermediate-scale quantum
(NISQ) regime\cite{preskill2018quantum,huang2023near}. Limited qubit counts, imperfect gate fidelities, and restricted
circuit depths make it difficult to execute many practically relevant quantum
algorithms directly on available hardware.

As quantum computing advances from NISQ devices toward early fault-tolerant
systems\cite{katabarwa2024early}, integrating quantum processors into classical high-performance
computing environments offers a viable path toward practical deployment.
A growing number of countries are pursuing this direction through research
programs and platform development, including initiatives at Oak Ridge National
Laboratory in the United States\cite{beck2024integrating}, RIKEN in Japan, and the Hefei
Quantum--Supercomputing Hybrid Computing Center in China.
These efforts point to an emerging
systems architecture that couples classical supercomputing systems with quantum
processors based on diverse hardware modalities\cite{elsharkawy2024integration}. A central systems challenge is
how to orchestrate task placement, data movement, communication, and execution
across such distributed quantum--classical resources\cite{shehata2026bridging}.

Quantum circuit cutting \cite{bravyi2016trading,piveteau2023circuit,peng2020simulating,chen2024quantum} exemplifies this challenge. It decomposes a large
quantum circuit into smaller subcircuits, executes them separately on available
quantum devices, and reconstructs the output distribution of the original
circuit through classical post-processing. By reducing the size of the circuits
executed on quantum hardware, circuit cutting alleviates qubit-capacity
limitations and lowers the hardware requirements of the quantum execution
stage\cite{ying2023experimental}. However, it also transfers substantial computation, communication, and
storage overhead to the classical reconstruction stage.

Representative studies on quantum circuit cutting have mainly
focused on cut selection, circuit decomposition, and reducing
quantum-side execution overhead\cite{bravyi2016trading,mitarai2021constructing,mitarai2021overhead,peng2020simulating}. These efforts improve
the feasibility of circuit cutting on near-term quantum devices. However, as
the circuit size, the number of cuts, and the number of measured subcircuit
output states increase, classical reconstruction itself can become a major
performance and storage bottleneck.

Existing post-processing methods broadly fall into several categories. One line
of work attempts to reconstruct the full \(2^n\)-dimensional output probability
vector from subcircuit measurement results. Although this dense reconstruction
target preserves the complete output space, its computation and storage costs
grow exponentially with the number of qubits, limiting its applicability to
relatively small circuits. Another line of work reduces the explored output
space through high-probability state search, sampling, or approximate
distribution estimation. Lowe et al.\cite{lowe2023fast} reduced the post-processing cost of circuit
cutting by introducing randomized measurements. Perlin et al.\cite{perlin2021quantum} developed
maximum-likelihood fragment tomography, which estimates a probability-constrained
output distribution from fragment measurement data and thereby reduces the
classical resource cost of circuit-cutting post-processing. Tang et al.\cite{tang2021cutqc} proposed CutQC, which performs full-state reconstruction for small-scale instances and, for larger instances, employs
dynamic-definition (DD) queries to merge output states into probability bins
and recursively refine high-probability regions, thereby reducing
the storage pressure caused by full state-space reconstruction. Chen et al.\cite{chen2022approximate} introduced an MCMC-based approximate
reconstruction method that samples high-probability output bit strings without
exhaustively evaluating the entire distribution. Following this sampling-based
line, Lian et al.\cite{lian2023fast} developed an HMC-based fast reconstruction algorithm (FRA) to
improve high-probability-state sampling efficiency and reduce circuit-cutting
post-processing overhead.
Recent work has explored how sparsity can be exploited to reduce the computational cost of quantum computation. In the context of circuit-cutting reconstruction, Li et al.\cite{li2024case} exploited topology and determinism in NISQ variational ansatzes to eliminate unnecessary subcircuit configurations and employed sparse tensor contraction to reduce reconstruction cost. Beyond circuit cutting, state sparsity has also been leveraged in classical quantum simulation to reduce memory consumption and simulation time \cite{jaques2022leveraging}. However, these approaches mainly focus on reducing the computational complexity or resource requirements of quantum computation, while the scalable execution and resource management of the resulting reconstruction workloads on distributed heterogeneous platforms remain insufficiently explored. To address this system-level gap, this paper focuses on the scalable execution of circuit-cutting reconstruction workloads generated from sparse state data on a distributed CPU+DCU heterogeneous platform, where DCU (Deep Computing Unit) accelerators are used as device-side computing resources for data-parallel reconstruction operations. Rather than explicitly maintaining a dense \(2^n\)-dimensional probability vector or returning only a small set of high-probability samples, the framework efficiently manages sparse state data generated from subcircuit measurement results and performs distributed aggregation of their probabilities.

Heterogeneous computing is widely used to improve performance and
scalability in large-scale scientific computing. Heterogeneous
programming models and runtime systems support performance-portable
execution across CPUs, GPUs, and other accelerators through
abstractions for parallel execution and memory management, together
with mechanisms for task scheduling, resource orchestration, and
data management
\cite{trott2021kokkos,kim2024iris}.
In such platforms, CPUs typically handle control and orchestration,
while suitable data-parallel kernels are offloaded to accelerators.
Distributed multi-GPU runtimes can further coordinate concurrent
streams and device resources to improve execution efficiency
\cite{zhang2022co}.
Efficient heterogeneous execution also depends on coordinated
optimization of computation, data movement, and communication
\cite{zhang2021petscsf,tan2021optimizing}, while memory-aware
sparse formats can improve GPU memory-access efficiency for sparse
workloads
\cite{karimi2022vcsr}.

Heterogeneous and distributed architectures have also been widely
adopted in classical quantum circuit simulation
\cite{ahmadzadeh2023fast,haner20175,wu2019full,
liu2021closing}.
Large-scale simulation systems have explored distributed-memory
and many-core execution, state-vector and density-matrix simulation,
unified programming models, and hierarchical multi-GPU partitioning
\cite{li2019quantum,jones2019quest,
zhang2022uniq,xu2024atlas}.
Tensor-network simulators have further employed scalable HPC
backends and GPU-optimized tensor contractions
\cite{nguyen2022tensor,pan2024efficient}.
Performance-modeling studies have examined the effects of qubit
count, circuit depth, memory capacity, CPU/GPU configuration,
resource heterogeneity, and load imbalance on hybrid distributed
quantum circuit simulation
\cite{ahmadzadeh2024performance}. Other work has focused on GPU-oriented execution and heterogeneous
memory management, including circuit-pattern-based region
partitioning, GPU-kernel selection, joint use of host and device
memory, and communication--computation pipelining
\cite{zhang2021hyquas,feng2021dgquest}.
Communication optimization has also received increasing attention
through communication--partition co-optimization, one-step
communication, and communication-aware circuit transformation and
slicing
\cite{jiao2026communication,song2023quanpath,
xu2026minimizing}.

Collectively, these studies demonstrate the utility of
heterogeneous and distributed computing for large-scale classical
quantum circuit simulation. However, they predominantly target
direct simulation and organize computation around simulator-state
representations such as state vectors, density matrices, or
tensors, together with the associated communication and memory
management. In contrast, circuit-cutting post-processing operates on sparse state data generated from subcircuit measurement results. This distinct workload calls for computation, communication, data movement, and storage management organized around sparse state data rather than around the direct evolution of simulator states.

This paper studies classical post-processing reconstruction for quantum circuit
cutting on a CPU+DCU heterogeneous supercomputing platform. To address the high computational cost, global state indices that exceed 64 bits, and limited memory capacity associated with large-scale circuit-cutting reconstruction over sparse state data, we present a heterogeneous
parallel framework for circuit-cutting post-processing on the Songshan
supercomputer. The framework exploits heterogeneous computing resources and
hierarchical storage to support scalable execution of reconstruction workloads. The main contributions of this paper are summarized as follows:

(1) We design a CPU+DCU heterogeneous parallel execution framework for
classical post-processing reconstruction in quantum circuit cutting on
distributed heterogeneous platforms. Experimental results demonstrate that
the framework substantially accelerates reconstruction while maintaining high
reconstruction fidelity. Specifically, it achieves up to \(259\times\)
speedup over the optimized serial baseline on linear-cluster states and up to
\(4\times\) speedup over the homogeneous CPU-parallel baseline on random
circuits.

(2) We employ a two-word integer representation, consisting of high and low
words, to support global state indices beyond 64 bits. We also design a
three-level cooperative storage mechanism spanning device memory, host memory,
and out-of-core storage to mitigate the memory pressure caused by the rapid
growth of intermediate states. Together, these techniques enable the proposed
framework to complete reconstruction tasks at the
hundred-qubit scale, thereby improving the scalability of post-processing reconstruction in quantum circuit cutting.

The rest of this paper is organized as follows. Section II reviews quantum circuit cutting and classical reconstruction. Section III describes the CPU+DCU heterogeneous reconstruction method. Section IV evaluates speedup, scalability, and reconstruction fidelity. Section V analyzes performance sources and scalability bottlenecks from the perspectives of integer indexing, phase-level overhead, reconstructed-state counts, hierarchical storage, and cutting schemes. Section VI concludes the paper and outlines future work.

\section{Quantum Circuit Cutting and Reconstruction}

Quantum circuit cutting decomposes a correlated global circuit into independently executable subcircuits by cutting selected wires\cite{peng2020simulating}. Measurement and state-preparation operations are inserted at the cut points so that the correlations needed for reconstruction are retained. The subcircuit measurement results and probabilities are then combined in a classical post-processing stage to recover the output distribution of the original circuit.

In a quantum circuit, any unitary operator can be expanded over an orthogonal matrix basis. After normalization, the Pauli matrices \(\{I,X,Y,Z\}/\sqrt{2}\) form a standard orthonormal basis. Hence, any \(2\times2\) matrix can be written as
\begin{equation}
A=\frac{\Tr(AI)I+\Tr(AX)X+\Tr(AY)Y+\Tr(AZ)Z}{2}.
\end{equation}
Expanding the Pauli matrices in terms of their eigenbasis decompositions gives
\begin{align}
I &= \proj{0}+\proj{1}, &
X &= \proj{+}-\proj{-}, \notag\\
Y &= \proj{+i}-\proj{-i}, &
Z &= \proj{0}-\proj{1}.
\end{align}
Substituting (2) into (1) and simplifying, we obtain
\begin{equation}
A=\frac{A_1+A_2+A_3+A_4}{2},
\end{equation}
where
\begin{align}
A_1 &= [\Tr(AI)+\Tr(AZ)]\proj{0}, \notag\\
A_2 &= [\Tr(AI)-\Tr(AZ)]\proj{1}, \notag\\
A_3 &= \Tr(AX)[2\proj{+}-\proj{0}-\proj{1}], \notag\\
A_4 &= \Tr(AY)[2\proj{+i}-\proj{0}-\proj{1}].
\end{align}
Because the projection measurement operators associated with the \(I\) and \(Z\) bases coincide in practical circuit execution, the subcircuit containing the quantum output terminal needs to be measured only in the \(X\), \(Y\), and \(Z\) bases.

\begin{figure}[!t]
\centering
\includegraphics[width=\linewidth]{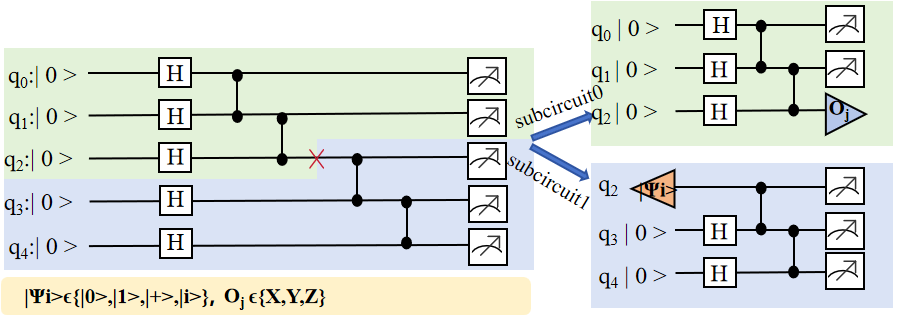}
\caption{Illustration of quantum circuit cutting.
\textit{Left:} A five-qubit circuit is cut on wire $q_2$ into two
independently executable three-qubit subcircuits.
\textit{Right:} At the cut boundary, subcircuit 1 measures the outgoing
qubit in the $X$, $Y$, and $Z$ bases, while subcircuit 2 prepares the
corresponding input state from
$\{|0\rangle, |1\rangle, |+\rangle, 
|+i\rangle\}$.
The output probability distribution of the original circuit is recovered
by classically combining the subcircuit results obtained under the
different measurement and state-preparation configurations.}
\label{fig:cutting}
\end{figure}

Fig.~\ref{fig:cutting} illustrates cutting and reconstruction on a 5-qubit example. A cut on \(q_2\) divides the original circuit into two three-qubit subcircuits, allowing a circuit that would otherwise require a five-qubit device to be executed as two smaller subcircuits. Consider the probability of the original output state \(\ket{10111}\). For the first subcircuit, the relevant output state is \(\ket{10}\). According to (4), its four reconstruction terms are
\begin{align}
p_{1,1}&=p(|100\rangle\!\mid\!I)+p(|101\rangle\!\mid\!I)+p(|100\rangle\!\mid\!Z)-p(|101\rangle\!\mid\!Z), \notag\\
p_{1,2}&=p(|100\rangle\!\mid\!I)+p(|101\rangle\!\mid\!I)-p(|100\rangle\!\mid\!Z)+p(|101\rangle\!\mid\!Z), \notag\\
p_{1,3}&=p(|100\rangle\!\mid\!X)-p(|101\rangle\!\mid\!X), \notag\\
p_{1,4}&=p(|100\rangle\!\mid\!Y)-p(|101\rangle\!\mid\!Y).
\end{align}
For the second subcircuit, the corresponding state is \(\ket{111}\), and the four terms can be written as
\begin{equation}
\begin{aligned}
p_{2,1}&=p(|111\rangle\!\mid\!|0\rangle),\quad
p_{2,2}=p(|111\rangle\!\mid\!|1\rangle),\\
p_{2,3}&=2p(|111\rangle\!\mid\!|+\rangle)-p(|111\rangle\!\mid\!|0\rangle)-p(|111\rangle\!\mid\!|1\rangle),\\
p_{2,4}&=2p(|111\rangle\!\mid\!|+i\rangle)-p(|111\rangle\!\mid\!|0\rangle)-p(|111\rangle\!\mid\!|1\rangle).
\end{aligned}
\end{equation}
The probability of the original circuit state is obtained by summing the four paired Kronecker products of the corresponding subcircuit outputs:
\begin{equation}
p(\ket{10111})=\frac{1}{2}\sum_{i=1}^{4}p_{1,i}\otimes p_{2,i}.
\end{equation}
We use the following fidelity metric to evaluate reconstruction quality:
\begin{equation}
\mathcal{F}=\left(\sum_s\sqrt{p_{\mathrm{th}}(s)p_{\mathrm{exp}}(s)}\right)^2,
\end{equation}
where \(p_{\mathrm{th}}(s)\) is the theoretical probability distribution and \(p_{\mathrm{exp}}(s)\) is the reconstructed distribution.

This example highlights the parallel structure of the reconstruction problem. For a single cut, reconstruction becomes a weighted sum of four independent Kronecker products. Within each Kronecker product, probability products for different state combinations are also independent. These properties expose substantial data parallelism in the post-processing stage.

For larger circuits, the cutting operation can be applied at multiple cut points, yielding several subcircuits that can be executed independently. If the number of cut edges is \(K\), the number of reconstruction terms grows as \(4^K\). Each term still requires subcircuit measurement results to be combined through Kronecker products, followed by accumulation of probabilities associated with identical global output states. Circuit cutting therefore reduces quantum-side hardware requirements, but it can shift a large amount of work to classical reconstruction. This classical burden becomes a limiting factor when the number of cuts or the number of nonzero-probability states in the reconstructed distribution grows.

\section{Heterogeneous Parallelization Method for Quantum Circuit Cutting Post-Processing}

\subsection{Heterogeneous Parallel Framework for Quantum Circuit Cutting Post-Processing}

Classical post-processing recombines subcircuit results obtained under different measurement bases and preparation configurations to recover the probability distribution of the original circuit. As the circuit size and cut count increase, reconstruction generates many intermediate states, and probability contributions corresponding to identical global states must be merged. A single process or a single hardware layer can therefore become bottlenecked by state expansion, probability computation, data exchange, and intermediate-state storage. We address these coupled bottlenecks with a multi-level cooperative framework for CPU+DCU heterogeneous platforms.

\begin{figure}[!t]
\centering
\includegraphics[width=\linewidth]{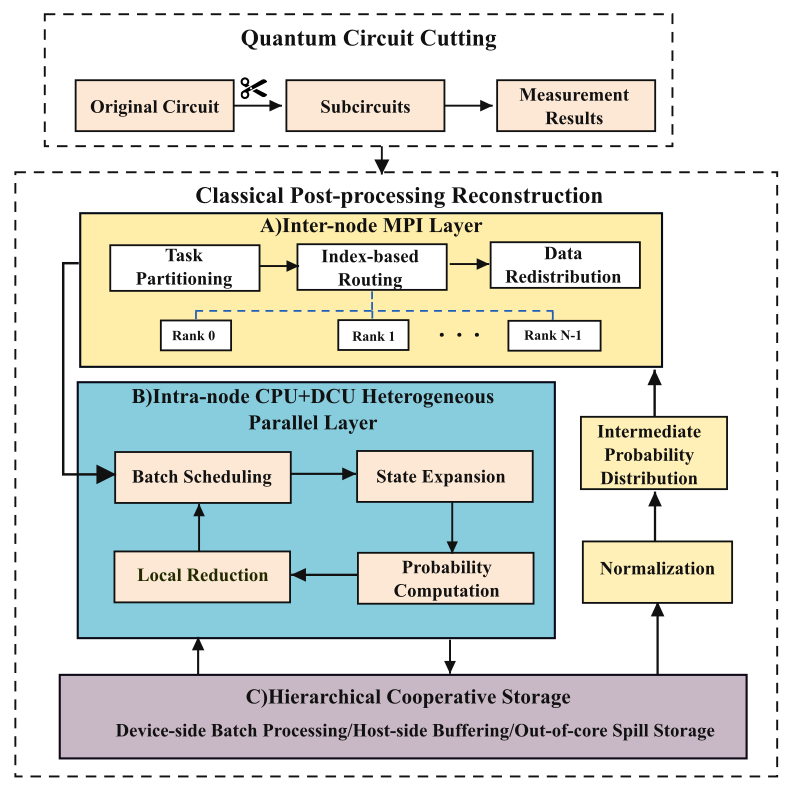}
\caption{Overall workflow of the proposed CPU+DCU heterogeneous parallel
framework for circuit-cutting post-processing reconstruction. The original
quantum circuit is first cut at selected locations into independently
executable subcircuits, which are executed on quantum devices to produce
measurement results for classical reconstruction. The inter-node MPI layer
performs sparse-state partitioning, assigns state indices to owner processes,
and redistributes newly generated states and their probability contributions
across processes. Within each node, CPUs and DCUs cooperate to perform classical
post-processing reconstruction, while device memory, host memory, and
out-of-core storage jointly manage intermediate sparse-state data.}
\label{fig:framework}
\end{figure}

As shown in Fig.~\ref{fig:framework}, the framework has three
layers: inter-node MPI cooperation, intra-node CPU+DCU
heterogeneous computation, and hierarchical cooperative storage.
At the inter-node layer, each probability contribution is assigned
to an owner process determined by the integer index of its global
basis state. Contributions with the same global state are routed
to the same MPI process and reduced locally. After all contributions
associated with the same global state index have been accumulated,
a global basis state \(s\) is retained only if its aggregated
reconstructed probability satisfies \(p(s)>0\); otherwise, it is
discarded. Throughout this paper, we refer to the retained states
as nonzero-probability states. Each process therefore maintains
only the states assigned to it, rather than a full copy of the
intermediate probability distribution. After each expansion round,
newly generated states are re-routed in batches according to their
integer indices, enabling distributed reduction without redundant
state residency across processes.

Within a node, the CPU and DCU perform complementary roles. The CPU organizes subcircuit outputs, constructs effective coefficients, schedules batches, manages communication buffers, performs local reductions, and controls out-of-core chunks. The DCU accelerates the regular part of the computation: batched state expansion and Kronecker-product probability updates. For a batch of input states and a current subcircuit nonzero-probability state set, expansion consists mainly of independent index concatenations and probability multiply-add operations, which map naturally to device-side data parallelism. The storage hierarchy is organized around data lifetime: short-lived expansion operands remain in device memory, communication-ready contributions are staged in host memory, and overflow buckets are spilled to out-of-core storage for later routing, communication, and state update.

The framework is organized around a unified global basis-state index. This index connects state generation, probability computation, owner-process selection, inter-process communication, and hierarchical storage into one dataflow. Each reconstruction round performs local expansion, probability update, owner-based routing, inter-process redistribution, and local reduction. The process continues until all subcircuit results have been incorporated into the reconstructed probability distribution. Section III-B describes the high/low-word indexing and routing scheme, and Section III-C presents the hierarchical cooperative storage mechanism.

\subsection{Quantum State Index Representation and Routing Strategy}

Intermediate quantum states repeatedly undergo expansion, caching, communication, and reduction during reconstruction. Their representation therefore affects not only local computation, but also data transmission, owner-process selection, and duplicate-state detection. String-based or high-level object representations require repeated format conversion, memory allocation, and serialization during concatenation and communication; these costs grow with the number of intermediate states. We instead use a unified integer representation for basis states. Bit strings are mapped to binary integers, and storage, transmission, expansion, routing, and merging are all organized around this index. State expansion is thereby reduced to shifts, bitwise combinations, and masks, which are well suited to batched CPU and DCU execution.

A single 64-bit integer is sufficient when the global state width does not exceed 64 bits. For wider states, we use a high/low-word representation: the low field stores the lower 64 bits, and the high field stores the remaining upper bits. Appending a local suffix index to an existing global index is implemented by shifting the current index by the local bit width, moving the overflow beyond the 64-bit boundary into the high field, inserting the local suffix into the low field, and applying a mask for the target width. This representation provides a uniform index format across circuit scales and enables contiguous arrays for input indices, expanded indices, and probability values during device-side batch execution.

Algorithm~\ref{alg:index} gives the expansion procedure. In the algorithm, $\textit{High}$ and $\textit{Low}$ denote the upper and lower 64-bit words of the current global state index, respectively; $w$ denotes the valid bit width of the current global index; $L$ is the list of local suffix indices produced by the next subcircuit; and $b$ denotes the bit width of each local suffix index. The expanded index width is $W=w+b$, and $\textit{ExpandedIndexList}$ stores the global indices obtained by appending each suffix in $L$ to the current index.

\begin{algorithm}[!t]
\caption{High/Low Integer Index Expansion}
\label{alg:index}
\begin{algorithmic}[1]
\REQUIRE Current global state index \((\mathit{High},\mathit{Low})\); list of local suffix indices \(L\); bit width \(b\) of each local suffix index, where \(0<b<64\); bit width \(w\) of the current global index.
\ENSURE \(\mathit{ExpandedIndexList}\), the list of expanded global state indices.
\STATE \(\mathit{ExpandedIndexList}\leftarrow\emptyset\)
\STATE \(W\leftarrow w+b\)
\STATE \(\mathit{oldHigh}\leftarrow \mathit{High}\)
\STATE \(\mathit{oldLow}\leftarrow \mathit{Low}\)
\FOR{each \(\mathit{localIndex}\in L\)}
\STATE \(\mathit{localIndex}\leftarrow \mathit{localIndex}\ \&\ (2^b-1)\)
\STATE \(\mathit{overflowBits}\leftarrow \mathit{oldLow}\gg(64-b)\)
\STATE \(\mathit{newLow}\leftarrow (\mathit{oldLow}\ll b)\ |\ \mathit{localIndex}\)
\STATE \(\mathit{newLow}\leftarrow \mathit{newLow}\ \&\ (2^{64}-1)\)
\IF{\(W>64\)}
\STATE \(\mathit{newHigh}\leftarrow (\mathit{oldHigh}\ll b)\ |\ \mathit{overflowBits}\)
\STATE \(\mathit{newHigh}\leftarrow \mathit{newHigh}\ \&\ (2^{W-64}-1)\)
\ELSE
\STATE \(\mathit{newHigh}\leftarrow0\)
\STATE \(\mathit{newLow}\leftarrow \mathit{newLow}\ \&\ (2^W-1)\)
\ENDIF
\STATE append \((\mathit{newHigh},\mathit{newLow})\) to \(\mathit{ExpandedIndexList}\)
\ENDFOR
\end{algorithmic}
\end{algorithm}

The unified integer index is used not only for local state expansion, but
also as the routing key for inter-process redistribution and duplicate-state
reduction. We adopt a deterministic owner-routing strategy based on
SplitMix64\cite{steele2014fast} to assign each intermediate state to an MPI process. Specifically,
the global state index is first transformed by a fixed 64-bit mixing function,
and the resulting hash value is mapped to an owner process by modulo reduction
over the number of MPI processes. For a state index whose width does not
exceed 64 bits, the routing function applies SplitMix64 directly to the low
64-bit word. For a wider state index, the high and low 64-bit words are mixed
separately and combined by bitwise XOR before the final modulo mapping. This deterministic
mapping ensures that probability contributions associated with the same global
basis state are always routed to the same owner process for accumulation. It
also reduces the sensitivity of load distribution to consecutive indices and
low-bit regularities.

After each expansion round, a newly generated state may no longer belong to the source process. The source process recomputes the owner of each new index and groups indices and probability contributions by target process. Batched communication then sends each group to the corresponding owner, where contributions with identical global indices are accumulated. This batched redistribution improves load balance when intermediate-state distributions are uneven while preserving deterministic ownership for duplicate-state reduction.

Because generation, routing, communication, out-of-core organization, and reduction all use the same integer key, the framework avoids repeated representation conversion and deterministic duplicate detection remains valid across MPI processes. This common indexing rule is what connects local device-side expansion to global distributed aggregation.

\subsection{Cooperative Storage Strategy}

\begin{figure*}[!t]
\centering
\includegraphics[width=0.90\textwidth]{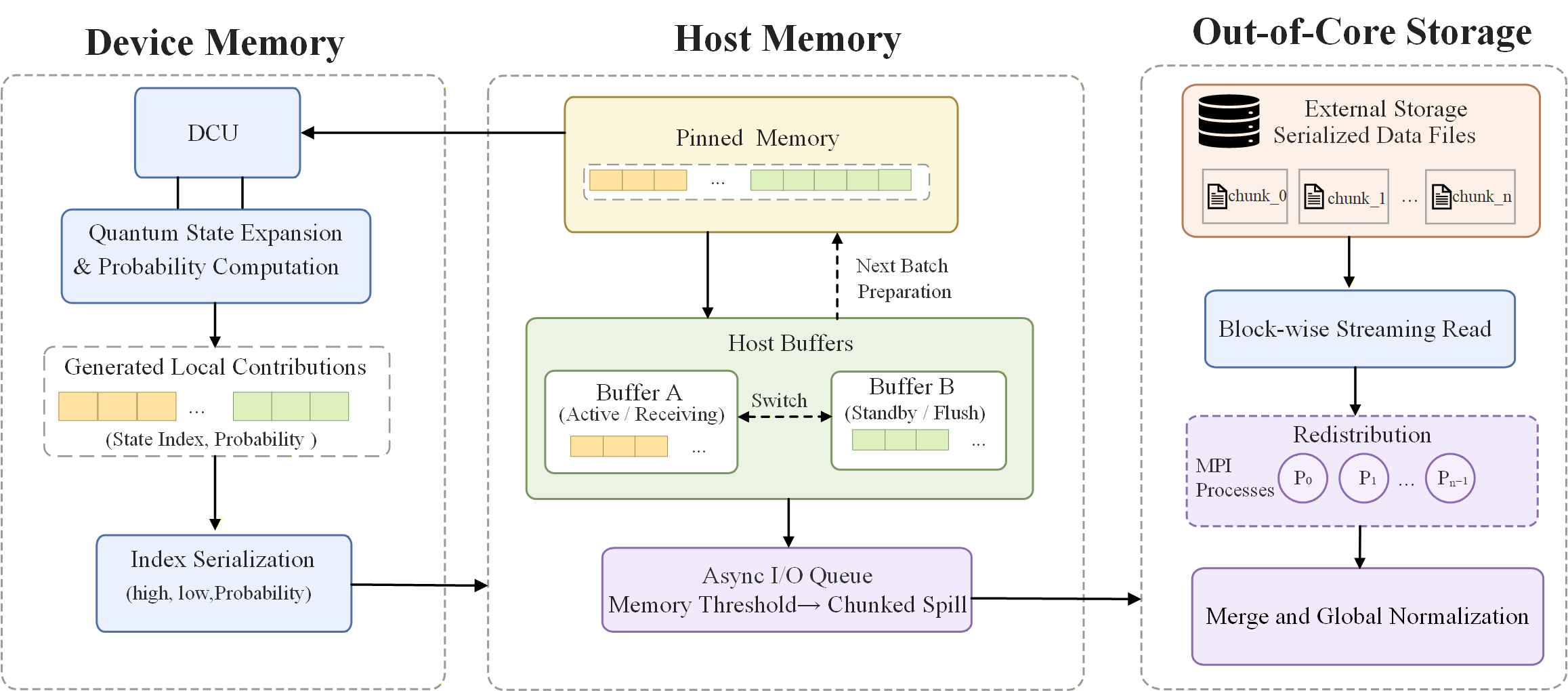}
\caption{Schematic of the hierarchical cooperative storage mechanism.
Device memory performs DCU-side state expansion and probability computation
and generates local state contributions for redistribution. Host memory uses
pinned memory, buffering, and asynchronous I/O queues for data staging, batch
preparation, and spill management. When host-memory usage exceeds a predefined
threshold, intermediate data are asynchronously written to disk and later
reintroduced into the processing pipeline through block-wise reading,
redistribution, and merging. The three storage levels cooperate through
buffered and asynchronous I/O pipelines, mitigating the impact of single-level
capacity limitations on the scalability of large-scale reconstruction.}
\label{fig:storage}
\vspace{-0.8em}
\end{figure*}

The number of intermediate states can exceed the capacity of either device memory or host memory as reconstruction proceeds. Although distributed routing spreads the global state set across MPI processes, each process may still produce large local bursts during expansion, redistribution, and reduction. The storage system therefore manages intermediate data according to lifetime and access pattern, as shown in Fig.~\ref{fig:storage}.

Device memory holds only the operands and output buffers required by the current DCU batch, including input indices, probability values, effective coefficients, and generated contributions. Once a batch finishes, the generated indices and probability contributions are returned to the host, and the device buffer is reused. Host memory acts as a staging and reduction layer: it receives device outputs, performs group caching and local merging, organizes buckets, and prepares communication buffers. Pinned memory and a multi-buffer pipeline overlap input preparation, device computation, result transfer, and host-side processing, reducing synchronization between the CPU and DCU stages.

When locally generated states exceed the host-memory cache threshold, the out-of-core path is activated\cite{wang2025towards}. The host computes the target process and bucket ID for each intermediate state from the global integer index and organizes contributions into group caches keyed by \((\textit{target process},\textit{bucket ID})\). The spill threshold bounds the cache size; once the threshold is exceeded, groups with larger state counts are written preferentially to out-of-core storage. The block-granularity parameter controls the size of write, read, and chunk-level merge operations, trading off I/O frequency, merge overhead, and per-block memory pressure. When asynchronous I/O is enabled, background write tasks reduce foreground stalls during computation and data organization\cite{park2022snuqs}.

Spilled intermediate data are stored as fixed-width binary arrays. Each state consists of a low word, an optional high word, and a fixed-width floating-point probability value. Compared with string-based state storage, this format reduces metadata overhead and improves sequential read/write efficiency. Compared with loading an entire spill file at once, the separated-array layout supports streaming reads, redistribution, and bucket-level merging at chunk granularity.

The hierarchy separates short-lived compute data from longer-lived communication and spill data. Device memory supplies high-throughput batched expansion, host memory stages and reduces communication-ready contributions, and out-of-core storage absorbs intermediate results that exceed the memory threshold. Because spilled data retain the same integer-index format, they can be read back and re-enter routing and probability updates without semantic conversion. The result is a reconstruction path that reduces memory pressure while preserving deterministic aggregation and reconstruction fidelity.

\section{Results}

\subsection{Experimental Setup}

We evaluate the CPU+DCU heterogeneous post-processing framework on four classes of quantum circuits with different output structures and reconstruction characteristics.

\textit{(1) Linear-cluster states.} Linear-cluster states \cite{nielsen2006cluster} are representative many-body entangled states. They are typically built by applying Hadamard gates to the initial qubits and adding controlled entangling gates along a one-dimensional chain. Their output distributions are relatively uniform, and the number of nonzero-probability states in the reconstructed distribution grows rapidly with circuit scale.

\textit{(2) GHZ.} GHZ states\cite{greenberger1989going} are typically prepared using one Hadamard gate followed by a cascade of CNOT gates. Their theoretical output distribution is concentrated on the all-zero and all-one basis states, with probability 0.5 assigned to each.

\textit{(3) Bernstein--Vazirani (BV).} The Bernstein--Vazirani algorithm\cite{bernstein1993quantum} solves the hidden-string problem and produces highly deterministic outputs. Its measurement distribution is concentrated on the target basis state determined by the hidden string, with theoretical probability 1.

\textit{(4) Random circuits.} Random circuits are composed of randomly generated single-qubit and two-qubit gates. They provide less structured workloads that emulate general circuit behavior in gate distribution, entanglement pattern, and output probability structure.

These benchmarks serve complementary purposes. Linear-cluster states evaluate performance and fidelity as the reconstructed distribution contains more nonzero-probability states. GHZ and BV test large-scale reconstruction fidelity on structured distributions. Random circuits evaluate the benefit of heterogeneous acceleration in less regular settings. The experimental platform is summarized in Table~\ref{tab:config}.

\begin{table}[H]
\centering
\caption{Experimental Parameter Configuration}
\label{tab:config}
\footnotesize
\begin{tabular}{ll}
\toprule
Parameter Category & Configuration \\
\midrule
CPU & Hygon C86 7185 \\
Memory Capacity & 128 GB \\
Accelerator & Z100 (4 devices) \\
Device Memory & 16 GB \\
Disk & 50 TB \\
OS & CentOS 7 \\
ROCm / DTK & DTK 23.04 \\
Python & 3.8.5 \\
PyTorch & 2.0.0a0 \\
mpi4py & 4.1.1 \\
\bottomrule
\end{tabular}
\vspace{-0.4em}
\end{table}

\subsection{Overall Speedup}

\begin{figure}[H]
\centering
\includegraphics[width=0.98\linewidth]{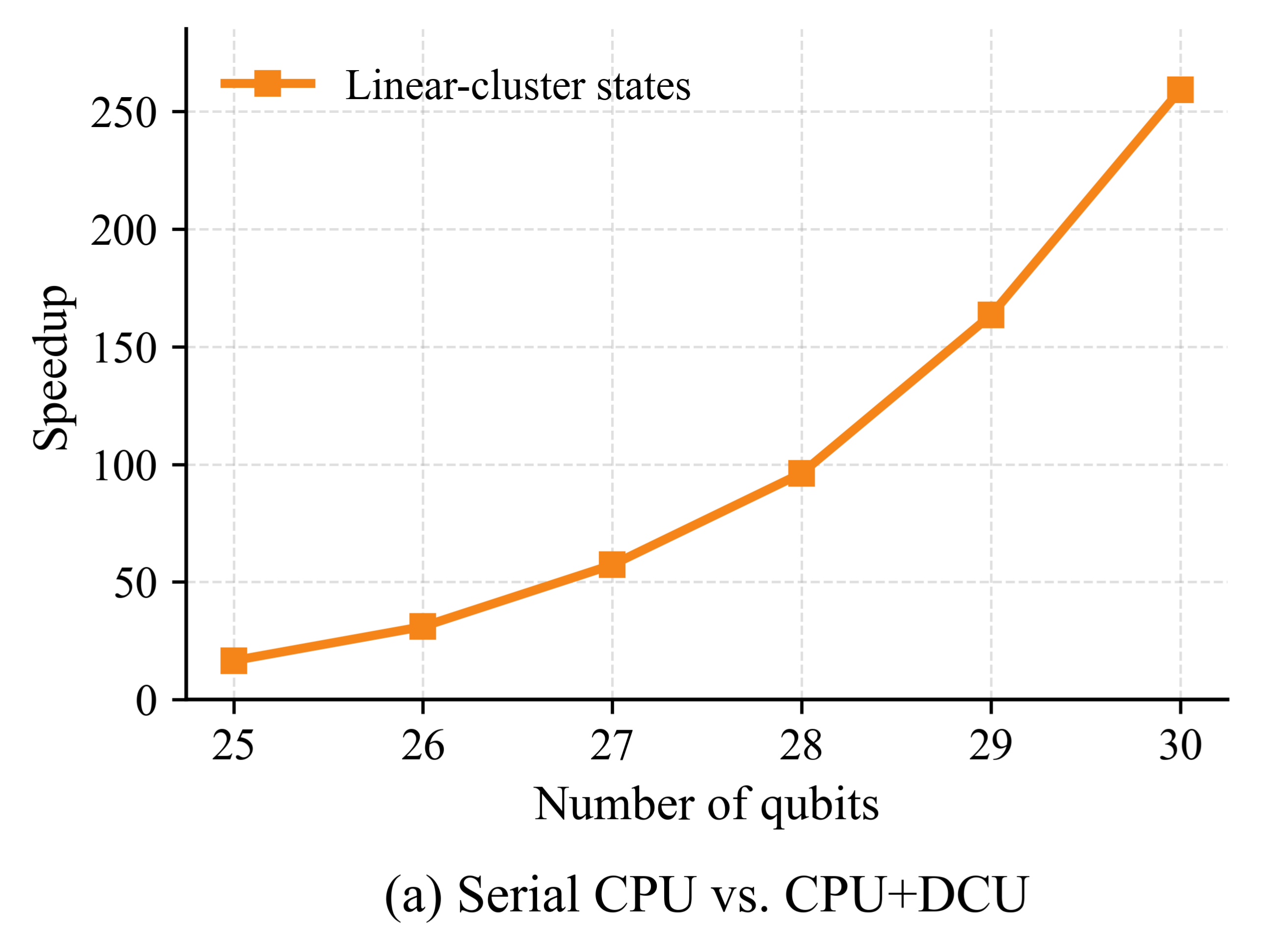}\\[-0.5ex]
\includegraphics[width=0.98\linewidth]{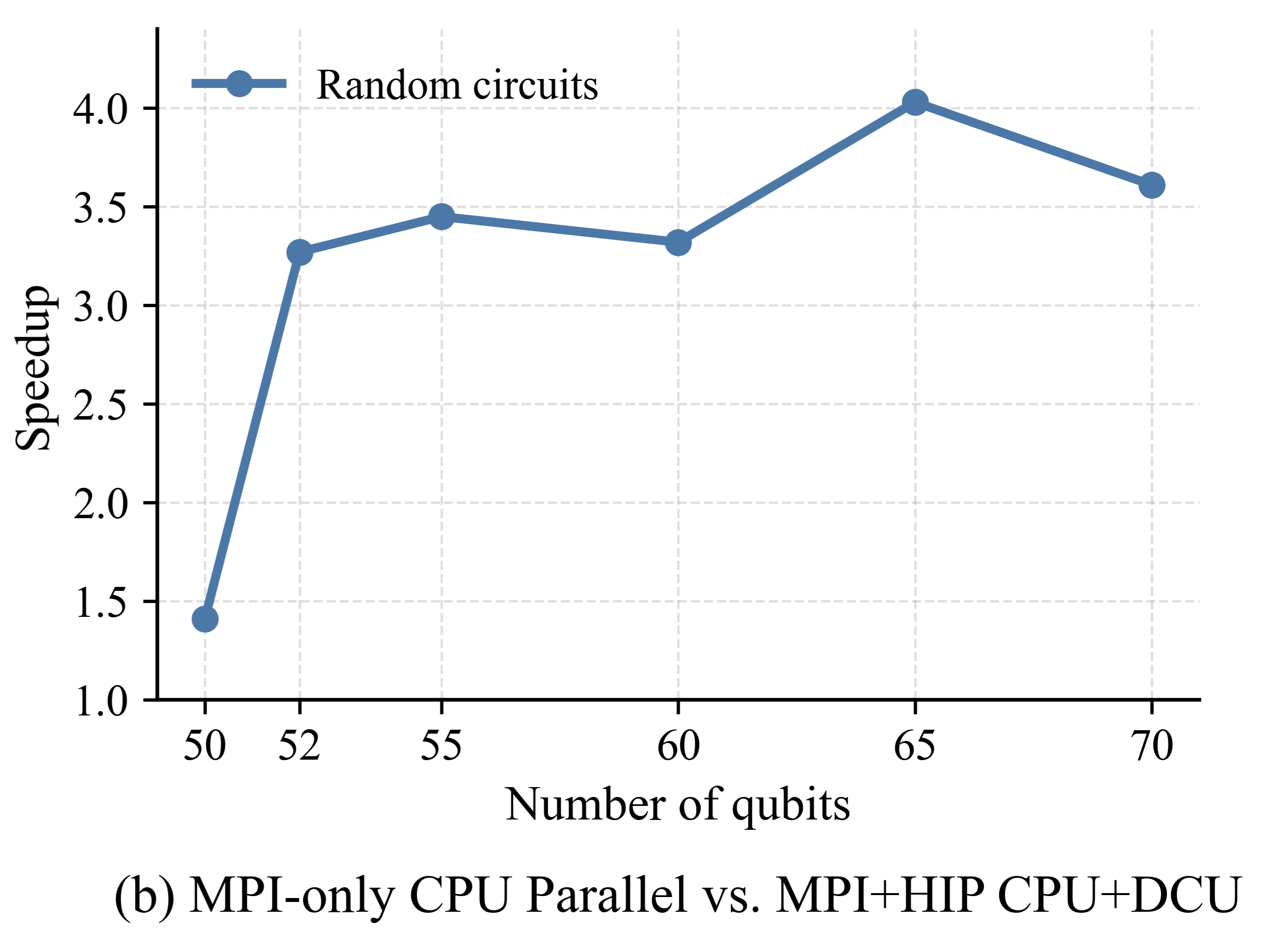}
\caption{Speedup evaluation of the CPU+DCU heterogeneous framework.
(a) Speedup over the optimized serial CPU implementation for linear-cluster states.
(b) Speedup over the homogeneous CPU-parallel implementation for random circuits.}
\label{fig:speedup}
\end{figure}

We compare three post-processing implementations: the optimized serial CPU implementation, the homogeneous CPU-parallel implementation, and the proposed CPU+DCU heterogeneous implementation. In Fig.~\ref{fig:speedup}(a), the optimized serial baseline runs with a single process on one node, whereas the CPU+DCU implementation uses up to five nodes, depending on the circuit scale. This comparison reports the end-to-end speedup of distributed heterogeneous reconstruction over the optimized serial baseline. In Fig.~\ref{fig:speedup}(b), the homogeneous CPU-parallel baseline and the CPU+DCU implementation use the same 60-node configuration, with DCU offloading being the only difference. This comparison isolates the additional benefit of device-side acceleration.

For linear-cluster states, the heterogeneous implementation obtains increasing speedup as the reconstruction workload grows, reaching approximately 259\(\times\) at 30 qubits. This behavior is expected because state expansion and probability accumulation in this benchmark are relatively regular and expose substantial data parallelism. As the scale increases, the CPU+DCU implementation can better amortize scheduling, communication, and device-transfer overheads, allowing distributed execution and DCU acceleration to contribute more effectively.

On random circuits, the heterogeneous implementation achieves approximately 1--4\(\times\) speedup over the homogeneous CPU-parallel baseline. The improvement is smaller and non-monotonic because random circuits produce different subcircuit output distributions, cutting positions, and intermediate-state distributions at different scales; the reconstruction cost is therefore not determined by qubit count alone. Nevertheless, the results show that DCU offloading remains beneficial even after CPU-side parallelization, especially for highly concurrent state expansion and probability-contribution computation.

\subsection{Scalability}

\begin{figure}[H]
\centering
\includegraphics[width=0.92\linewidth]{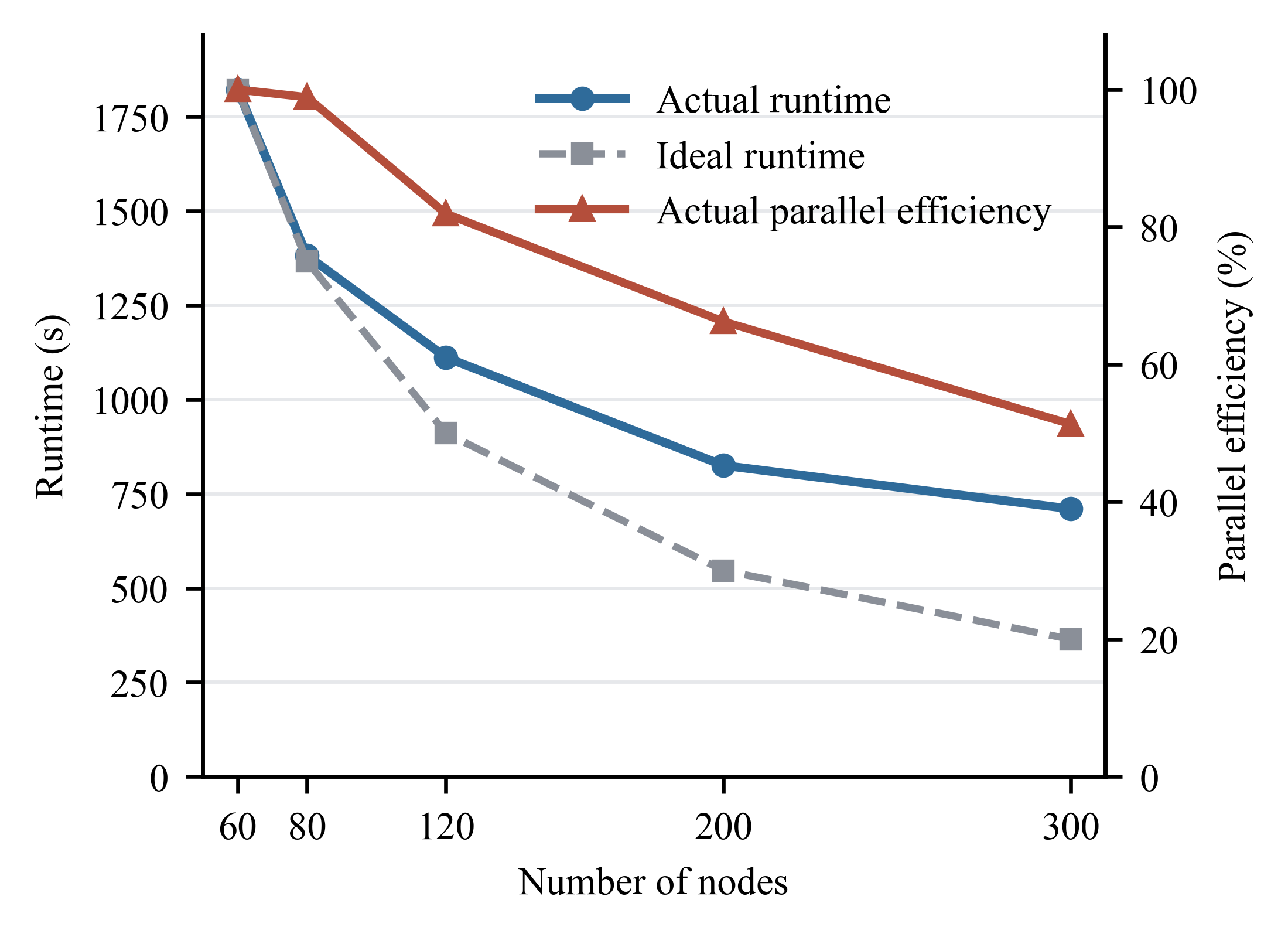}
\caption{Strong-scaling results for the post-processing reconstruction of a
fixed 53-qubit random circuit. Actual runtime denotes the measured runtime, while Ideal runtime
denotes the theoretical strong-scaling runtime calculated using the 60-node
configuration as the baseline. Taking the overall experimental feasibility
into account, the 60-node configuration is selected as the starting point and
baseline for the strong-scaling study.}
\label{fig:strong}
\end{figure}

Fig.~\ref{fig:strong} reports strong scaling on a fixed 53-qubit random-circuit
reconstruction problem. As the node
count increases from 60 to 300, the measured runtime decreases from 1821.90~s to
710.07~s. The Ideal curve uses the 60-node run as the baseline, and the
parallel efficiency at 300 nodes is 51.32\%.

The scaling curve shows that reconstruction becomes increasingly
communication- and I/O-limited as the node count grows. Increasing the
node count reduces the local expansion and probability-computation work
assigned to each process, but it does not proportionally reduce owner-based
state redistribution, cross-process data movement, out-of-core chunking, or
intermediate-state merging. Notably, as the node count scales from 60 to 300, the computation share drops from 44.10\% to 35.59\%, while the combined communication and I/O share rises from 55.90\% to 64.41\%. The loss in strong-scaling
efficiency is therefore mainly caused by the shrinking compute fraction and
the growing relative cost of global data movement and intermediate-state
management. This result indicates that further scaling is limited less by
device-side arithmetic throughput alone than by communication scheduling,
state redistribution, and out-of-core state management.

\subsection{Fidelity Evaluation}

\begin{figure}[H]
\centering
\includegraphics[width=0.98\linewidth]{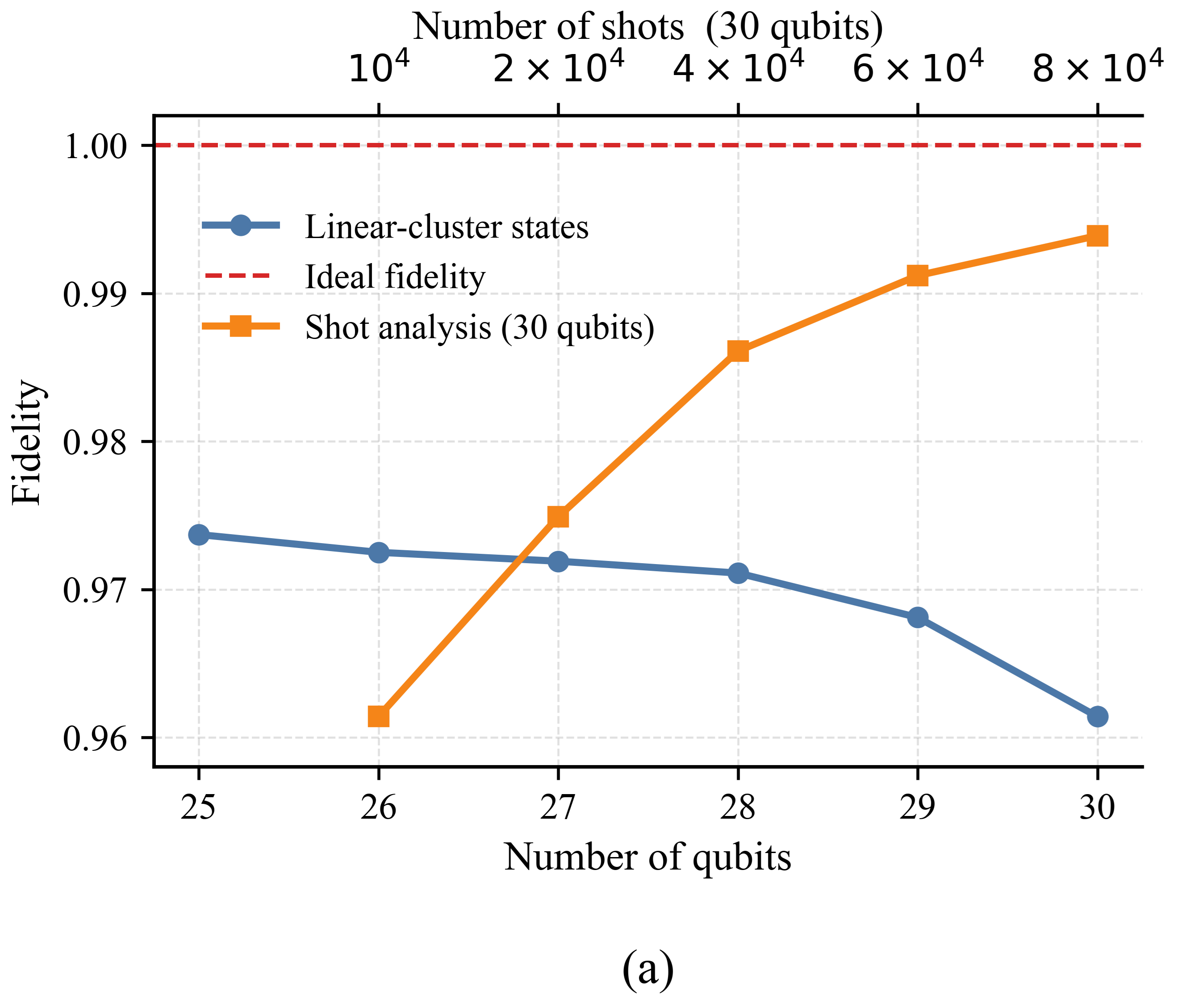}\\[-0.5ex]
\includegraphics[width=0.98\linewidth]{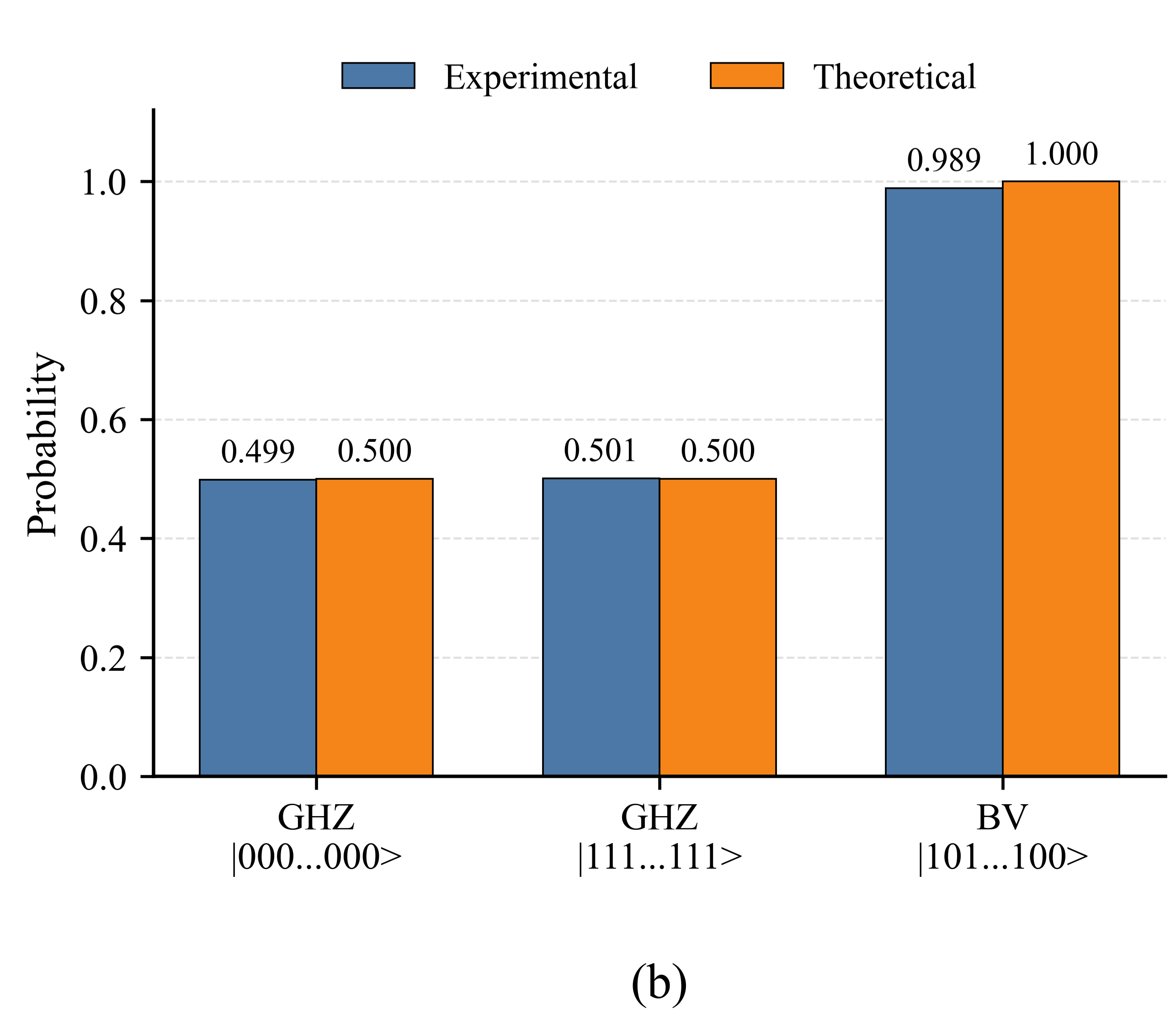}
\caption{Fidelity evaluation and large-scale structured-circuit validation.
(a) Fidelity evaluation of linear-cluster states with different qubit scales and the impact of measurement shot counts on the 30-qubit case. (b) Reconstructed probabilities of principal states for 118-qubit GHZ and BV.}
\label{fig:fidelity}
\end{figure}

For small-scale cases, we compute fidelity for linear-cluster states at different scales. For large-scale structured cases, we use 118-qubit GHZ and BV circuits and compare the reconstructed probabilities of their principal states with the corresponding theoretical distributions. Fig.~\ref{fig:fidelity} summarizes the results.

As shown in Fig.~\ref{fig:fidelity}(a), linear-cluster states maintain high reconstruction fidelity overall. Their relatively uniform output distributions, however, spread a fixed shot budget over more states as the circuit scale increases, so fidelity can decrease with scale. For the 30-qubit case, increasing the shot count drives fidelity toward the theoretical value, suggesting that the observed error is dominated by finite-sampling fluctuation rather than systematic bias in the reconstruction algorithm. In Fig.~\ref{fig:fidelity}(b), the reconstructed probabilities of the two principal GHZ states are close to 0.5, and the reconstructed probability of the BV target state is close to 1.0. These results indicate that the framework preserves the expected probability structure on the evaluated large-scale circuits.

\FloatBarrier
\section{Discussion}

We next analyze the sources of performance improvement and the factors limiting scalability. The discussion focuses on six aspects: state representation, runtime composition, reconstructed-state counts, hierarchical cooperative storage, communication strategy, and cutting schemes. We first evaluate whether integer-based indexing reduces the baseline overhead of reconstruction. We then decompose runtime into computation, communication, and I/O to identify how bottlenecks shift with scale. Next, we compare reconstructed-state counts across different methods to analyze storage pressure. Finally, we examine how hierarchical storage, communication strategy, and cutting schemes affect the scalability and cost of reconstruction under fixed resources.

\subsection{Impact of Integer Index Representation}

\begin{figure}[H]
\centering
\includegraphics[width=0.95\linewidth]{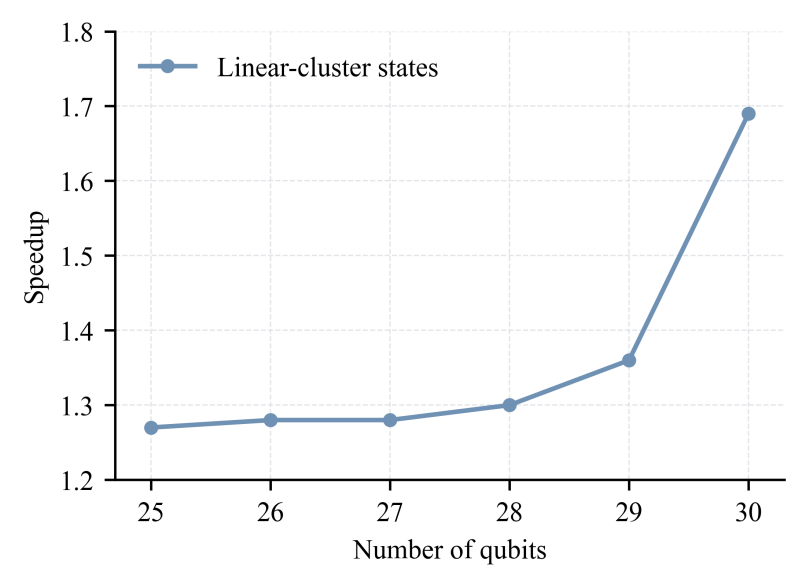}
\caption{Effect of quantum state integer indexing on serial post-processing speedup.}
\label{fig:index_speedup}
\end{figure}

To isolate the effect of state representation, we compare serial post-processing runtime before and after replacing string-based state handling with integer indexing on linear-cluster states. Fig.~\ref{fig:index_speedup} reports the resulting speedup.

The improvement is not merely a consequence of adding parallel resources. Integer indexing reduces the base cost of state concatenation, object management, and duplicate-state lookup. As the number of states grows, these representation-level costs become increasingly important, making integer indexing more beneficial at larger scales.

\subsection{Runtime Breakdown of Computation, Communication, and I/O}

To analyze the runtime composition of reconstruction, we fix the number of
reconstruction rounds to four and run 50--54-qubit random circuits on 60 nodes. Table~\ref{tab:breakdown} decomposes the runtime
into computation, communication, and I/O phases.

\begin{table}[H]
\centering
\caption{Runtime phase breakdown for random circuits of different scales}
\label{tab:breakdown}
\scriptsize
\begin{tabular}{ccccc}
\toprule
Qubits & Nonzero-probability states & Compute & Communication & I/O \\
\midrule
50 & $19.99\times10^{9}$  & 66.80\% & 14.65\% & 18.55\% \\
51 & $37.44\times10^{9}$  & 48.42\% & 21.21\% & 30.37\% \\
52 & $80.16\times10^{9}$  & 40.79\% & 23.06\% & 36.15\% \\
53 & $164.22\times10^{9}$ & 41.53\% & 24.79\% & 33.68\% \\
54 & $167.51\times10^{9}$ & 41.71\% & 26.39\% & 31.89\% \\
\bottomrule
\end{tabular}
\end{table}

Table~\ref{tab:breakdown} shows that the number of nonzero-probability states in the reconstructed probability
distribution for the evaluated random-circuit instances increases from
\(19.99\times10^9\) to \(167.51\times10^9\). As the reconstructed probability
distribution grows, the runtime composition changes substantially: the
computation share drops from 66.80\% to 41.71\%, while the combined
communication and I/O share increases from 33.20\% to 58.28\%. Starting from
the 51-qubit case, communication and I/O together account for a larger fraction
of the runtime than computation. These results suggest that, at larger
reconstruction scales, the end-to-end cost is increasingly governed by state
redistribution, cross-process aggregation, out-of-core spilling, and
block-wise intermediate-result management, rather than by local state
expansion alone.

\subsection{Comparison of Reconstructed-State Counts across Different Reconstruction Methods}

\begin{figure}[H]
\centering
\includegraphics[width=\linewidth]{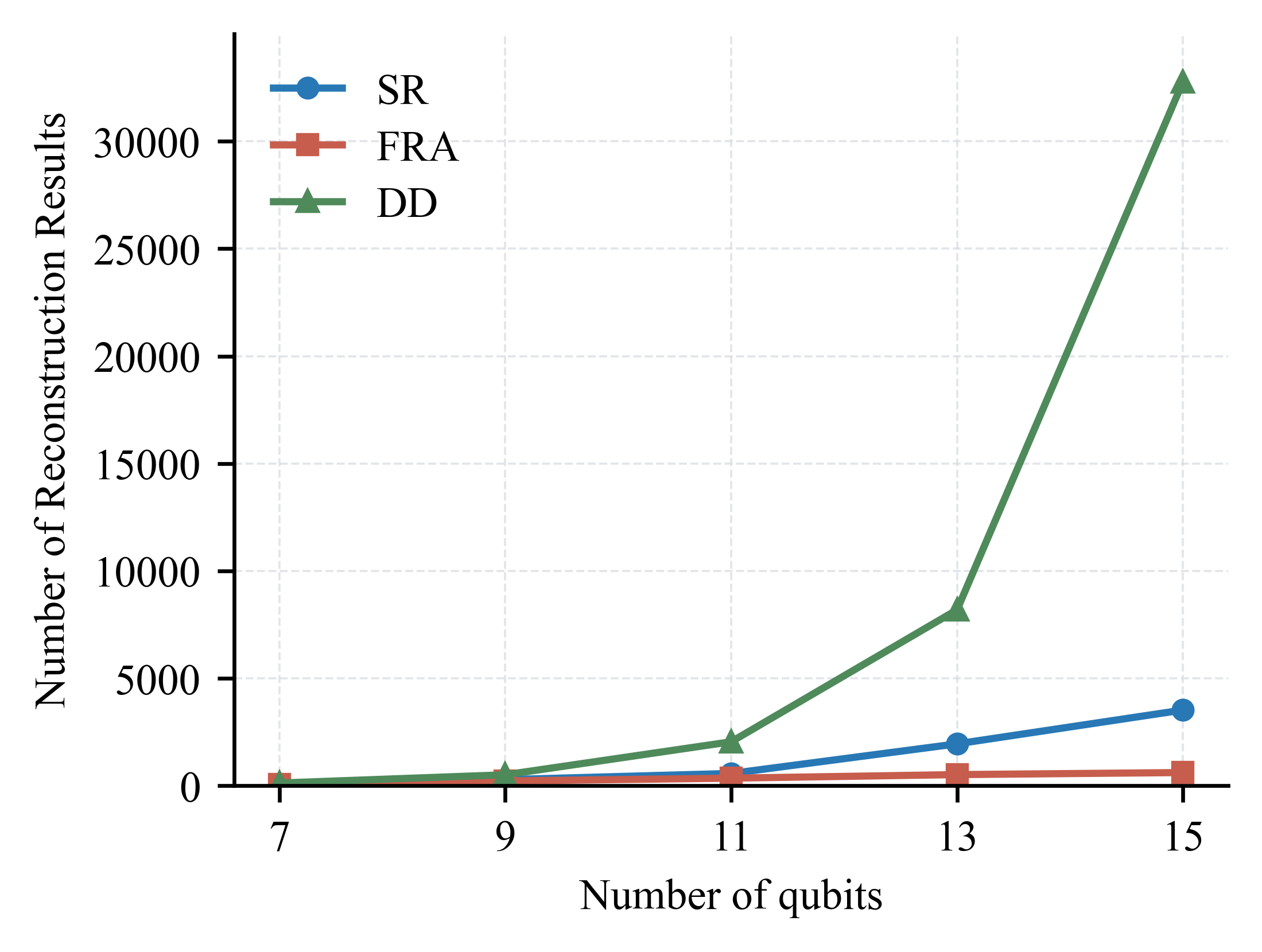}
\caption{Comparison of reconstructed-state counts for DD, FRA, and SR on
random circuits with different numbers of qubits.}
\label{fig:state_count_comparison}
\end{figure}

To further examine the storage pressure introduced by different reconstruction
strategies, we compare the number of reconstructed states produced by DD\cite{tang2021cutqc}, FRA\cite{lian2023fast},
and the sparse state execution strategy used in our framework, denoted as SR, under increasing qubit counts. The results are shown in
Fig.~\ref{fig:state_count_comparison}.

DD performs full-state reconstruction on the small-scale circuits. As the
circuit size increases, the number of states reconstructed by DD grows
exponentially, reaching \(32768\) states at 15 qubits. This result indicates
that full-state reconstruction can introduce substantial memory pressure as the
circuit size increases and the output space expands exponentially.
SR produces considerably fewer states because it keeps only output states whose
probabilities are nonzero in the distribution reconstructed from subcircuit
measurement results, rather than explicitly constructing a complete
\(2^n\)-dimensional probability vector. This reduces the number of states that must be stored, communicated, and merged during post-processing.
FRA yields the smallest state count because it relies on sampling and focuses
on high-probability solutions. This reduction, however, comes at the cost of
approximate reconstruction. SR is designed to recover the set of nonzero-probability
states, rather than only a sampled high-probability subset. Therefore, SR
retains more states than FRA while avoiding the full-state expansion required
by DD.

Although SR reduces storage overhead by avoiding direct full-state
reconstruction, the reconstruction process may still generate large-scale
intermediate data as the circuit size and complexity increase. For the proposed
framework, reducing the number of reconstructed states alone is therefore
insufficient to fully address the storage challenge in large-scale
post-processing. A hierarchical storage mechanism for intermediate data is also
needed to support reconstruction at larger scales.

\subsection{Effect of Hierarchical Cooperative Storage}

To evaluate the effect of the storage hierarchy under memory pressure, we use
the same 104-qubit random-circuit reconstruction instance as a controlled
storage-stress benchmark. The experiment is conducted on 10 compute nodes, with four MPI processes per node. Both execution modes use the same input circuit,
cutting scheme, reconstruction order, and runtime configuration; the only
difference is whether out-of-core spilling is enabled. This setting isolates
the contribution of hierarchical cooperative storage under a fixed hardware
budget. Table~\ref{tab:storage} reports the results.

\begin{table}[H]
\centering
\caption{Storage-stress results for the same 104-qubit random-circuit instance. Host-memory usage is measured at the MPI-rank level. For each MPI rank, we record the maximum resident set size observed during reconstruction. We then report the largest of these per-rank peak values across all MPI ranks.}
\label{tab:storage}
\begin{tabular}{lccc}
\toprule
Mode & Qubits & Peak memory/rank & Status \\
\midrule
In-memory only & 104 & 34.92 GiB & Failed \\
Hierarchical storage & 104 & 27.28 GiB & Completed \\
\bottomrule
\end{tabular}
\end{table}

The in-memory-only execution fails on this instance because the intermediate
working set exceeds the available host-memory budget. The reported
34.92~GiB per rank corresponds to the peak memory footprint observed before
failure. In contrast, with hierarchical storage enabled, including out-of-core
spilling, the same 104-qubit reconstruction instance completes successfully
with the maximum observed per-rank memory footprint being 27.28~GiB. This result shows that the
storage hierarchy improves completion capability under the same hardware
budget by moving spillable intermediate buckets from host memory to out-of-core
storage, rather than keeping the entire intermediate working set resident in
memory.

The result also demonstrates that hierarchical cooperative
storage can convert a memory-failing random-circuit reconstruction into a
completed run under fixed resources. At the same time, out-of-core spilling
introduces additional I/O operations, so storage management remains an
important scalability factor for larger reconstruction workloads.
\subsection{Analysis of the Communication Strategy}

In the proposed reconstruction workflow, communication volumes and message sizes are nonuniform because the number of intermediate contributions varies across processes and reconstruction rounds. To support variable message sizes without concentrating
intermediate data on a root process, we adopt a decentralized owner-based
redistribution scheme. Each global state index is assigned to a unique owner
rank, where all probability contributions associated with that state are merged
locally. This design distributes aggregation across MPI processes and avoids the
bandwidth and memory bottlenecks of root-centric aggregation.

During initialization, reconstruction metadata are broadcast, and the initial
sparse states are distributed according to their owner assignments using
\texttt{MPI\_Scatterv}. In each reconstruction round,
\texttt{MPI\_Alltoall} first exchanges the numbers of records destined for each
rank, after which \texttt{MPI\_Alltoallv} transfers the corresponding state
indices, probability values, and high-order index words when required. Each
owner rank then locally merges contributions with identical indices.
\texttt{MPI\_Allreduce} is finally used to compute global statistics required
for numerical reduction and normalization. Although skewed ownership may still
cause load imbalance, this communication scheme supports nonuniform message
sizes while keeping redistribution and aggregation fully distributed.
\subsection{Impact of Circuit Cutting Schemes}

Cutting schemes affect not only quantum-side feasibility but
also the shape of the classical reconstruction workload. Using
a 36-qubit linear-cluster state, we fix the number of
reconstruction rounds to four and evaluate the feasibility of
several cutting schemes without out-of-core storage. Table~\ref{tab:cuts} reports the results. In this table, columns Cir0--Cir4 denote
the five subcircuits produced by a cutting scheme, and each
entry gives the number of qubits assigned to the corresponding
subcircuit. The Success column indicates whether the classical
post-processing reconstruction can be completed under the
tested in-memory setting.

\begin{table}[H]
\centering
\caption{Feasibility Comparison under Different Cutting Schemes}
\label{tab:cuts}
\begin{tabular}{cccccc}
\toprule
Cir0 & Cir1 & Cir2 & Cir3 & Cir4 & Success \\
\midrule
9 & 9 & 9 & 9 & 4 & Yes \\
8 & 9 & 9 & 9 & 5 & Yes \\
10 & 6 & 9 & 9 & 6 & Yes \\
6 & 8 & 9 & 10 & 7 & No \\
11 & 7 & 7 & 7 & 8 & Yes \\
4 & 9 & 9 & 9 & 9 & No \\
\bottomrule
\end{tabular}
\end{table}

\begin{figure}[!t]
\centering
\includegraphics[width=\linewidth]{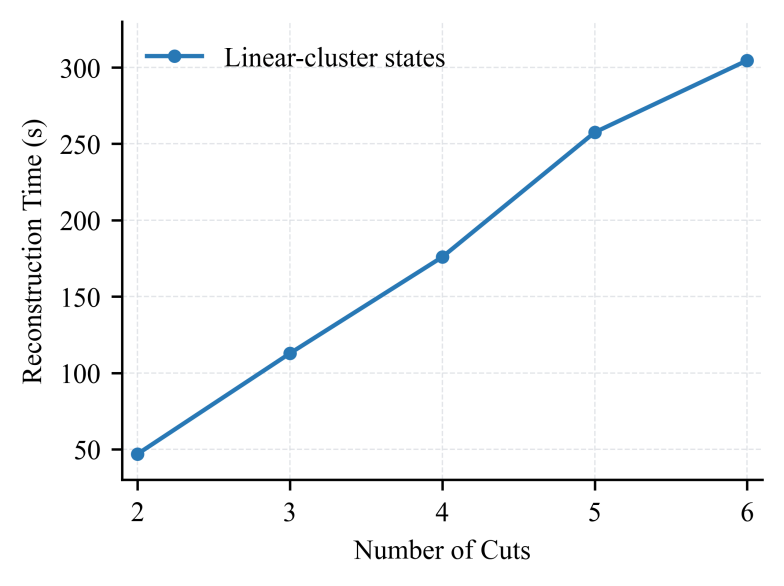}
\caption{Runtime variation of classical post-processing under different numbers of cuts.}
\label{fig:cuts}
\end{figure}

We also keep the circuit scale fixed and vary the number of cuts to measure how reconstruction time changes with the number of reconstruction rounds. Fig.~\ref{fig:cuts} shows the results.

Table~\ref{tab:cuts} shows that a cutting scheme that is feasible on the quantum side may still fail during classical reconstruction. The reason is that reconstruction feasibility depends on the order and size of intermediate nonzero-probability state sets. Similar to the greedy-subcircuit-order technique proposed by Tang et al.\cite{tang2021cutqc}, which reduces reconstruction overhead by optimizing subcircuit ordering, the reconstruction order also affects the growth of intermediate states in large-scale post-processing. If large subcircuits enter early reconstruction stages, many intermediate nonzero-probability states may be generated before enough reduction has occurred, exhausting memory. Introducing smaller subcircuits earlier can delay state growth and improve the likelihood of completion. Cutting-scheme design should therefore consider subcircuit size, access order, and combination path, not only the number of qubits assigned to each subcircuit.

Fig.~\ref{fig:cuts} further shows that reconstruction time increases with the number of cuts. More cuts reduce subcircuit size and ease quantum-side execution, but they also introduce more subcircuit results, more reconstruction terms, and more rounds of classical combination. Fewer cuts can reduce post-processing time, but they leave larger subcircuits that may require more quantum executions to obtain stable distributions and reduce statistical error. Practical circuit-cutting schemes must therefore balance quantum-device constraints against classical computation, communication, I/O, and storage cost.

We do not include a weak-scaling study because the actual workload of
circuit-cutting post-processing reconstruction is not determined by the qubit
count alone. It also depends on the circuit structure, entanglement pattern,
cutting scheme, and subcircuit measurement outcomes. Ensuring comparable
per-process workloads across different circuit scales is therefore difficult.
Instead, we use strong-scaling experiments and fixed-resource breakdowns by
execution phase to characterize parallel efficiency and the evolution of
performance bottlenecks.

The experiments are conducted using noiseless subcircuit measurement
data and are intended to isolate the numerical accuracy of the classical
post-processing reconstruction, rather than to evaluate end-to-end performance
on noisy quantum hardware. In practical settings, gate and readout errors,
together with finite-shot fluctuations, may perturb the estimated subcircuit
probabilities and consequently degrade reconstruction fidelity
\cite{ayral2021quantum}. Future work will therefore investigate reconstruction
robustness under realistic noise conditions and incorporate error-mitigation
and noise-aware techniques to reduce the impact of measurement errors
\cite{temme2017error}.

\FloatBarrier
\section{Conclusion and Future Work}

This paper presented a CPU+DCU heterogeneous framework for circuit-cutting post-processing reconstruction. The framework treats reconstruction as an end-to-end distributed dataflow over nonzero-probability states and combines heterogeneous CPU+DCU execution, high/low-word integer indexing, and hierarchical cooperative storage. Rather than constructing dense
\(2^n\)-state probability vectors, the framework reconstructs and manages the nonzero-probability states generated from subcircuit measurement results. 

The experimental results show that heterogeneous execution improves
post-processing performance while maintaining reconstruction fidelity. The strong-scaling results further show that reconstruction
becomes increasingly limited by communication scheduling, state redistribution,
and out-of-core state management as the node count grows. 

The reconstructed-state-count comparison shows that exploiting sparse state
data substantially reduces the number of states that must be stored compared
with full-state reconstruction, thereby alleviating storage pressure. The
storage experiment further demonstrates that hierarchical cooperative storage
improves the completion capability of reconstruction tasks under memory
constraints. Together with the large-scale GHZ and BV validation results, these experiments
show that the proposed framework can complete 
hundred-qubit-scale reconstruction tasks and improve the scalability of
classical post-processing reconstruction.

The results also show that cutting schemes directly affect the cost and
feasibility of classical post-processing. At a fixed circuit scale, different
schemes alter subcircuit sizes, reconstruction paths, and the number of
intermediate states generated during reconstruction. A cutting scheme that is
feasible on the quantum side may still fail during classical reconstruction if
it generates an excessively large number of intermediate states and exhausts
the available memory. Practical circuit-cutting design should therefore account
for both quantum-device constraints and the cost and feasibility of classical
post-processing.

Overall, the practicality of a large circuit-cutting task depends not only on
whether its subcircuits fit the available quantum hardware, but also on whether
the reconstruction workload induced by cutting can be completed efficiently
within the available classical resources. The current implementation is
limited to scenarios in which each subcircuit involves at most two cut points. Future work will extend the framework to more complex cutting scenarios, investigate reconstruction under realistic noise conditions, and incorporate error-mitigation and noise-aware techniques to improve reconstruction robustness.

In addition, we plan to port and evaluate the framework on the Sugon 8000
(Dengfeng) AI supercluster, which integrates a scaleFabric InfiniBand-class
native RDMA interconnect and the ParaStor distributed storage system.
Leveraging its high-speed interconnect and distributed storage capabilities, we will further optimize communication and out-of-core data access to improve the overall performance and scalability of large-scale reconstruction.

\section*{Data and Code Availability}

The source code and related data for implementing the proposed CPU+DCU heterogeneous parallel reconstruction framework are publicly available at https://github.com/Sunshine-tu/Cutqc-heterogeneous-reconstruction. The repository will remain publicly accessible throughout the review period.

\section*{Acknowledgments}

This work is supported by the National Key Research and Development Program of China (2024YFB4504103). This work is also supported by Jiangsu Province Engineering Research Center of IntelliSense Technology and System.


\end{document}